\documentclass[11pt]{article}

\usepackage{stix2}
\usepackage{pdfunicodemath}

\usepackage{booktabs,colortbl,enumitem,geometry,microtype,natbib,setspace,url,xcolor,hyperref,cleveref}
\crefformat{equation}{#2(#1)#3}
\crefname{figure}{figure}{figures}
\Crefname{figure}{Figure}{Figures}
\usepackage{caption}
\usepackage{pgfplots}
\usepgfplotslibrary{groupplots}
\pgfplotsset{compat=1.18}
\definecolor{qd2}{HTML}{5598E7}
\definecolor{qd3}{HTML}{2A78D6}
\definecolor{qd4}{HTML}{1C5CAB}
\definecolor{qd5}{HTML}{0D366B}
\pgfplotsset{
  quad/.style={
    width=\linewidth, height=0.8\linewidth,
    tick label style={font=\scriptsize}, label style={font=\footnotesize}, title style={font=\footnotesize, yshift=-1ex},
    axis line style={gray!60}, tick style={gray!60}, tick align=outside, tickpos=left,
    grid=major, major grid style={gray!18, line width=0.3pt},
    legend style={font=\footnotesize, draw=none, fill=none, /tikz/every even column/.append style={column sep=1.2em}},
  },
}

\hypersetup{colorlinks=true,citecolor=blue,urlcolor=blue,linkcolor=blue,
  pdftitle={Positive weight Hermite and Legendre quadrature rules},
  pdfauthor={Joris Pinkse},
  pdfsubject={Positive-weight Gauss-Hermite and Gauss-Legendre cubature rules for d = 2 to 5},
  pdfkeywords={cubature, quadrature, Gauss-Hermite, Gauss-Legendre, positive weights, node elimination, Moller bound}}
\setcitestyle{round}
\defcitealias{diablo}{DW}
\newcommand\headlineresults{This paper contains new 80-digit positive-weight Gauss-Hermite and positive-weight interior-node Gauss-Legendre quadrature rules for up to five dimensions and varying polynomial degree accuracy (depending on quadrature type and dimension).  Some of these rules improve on the best available rules in the literature and some offer rules where none (other than the tensor product) existed. The results were produced by combining methodology developed by previous researchers with new approaches.    The full precision rules themselves are at \url{https://doi.org/10.5281/zenodo.22881864}.  Software in Julia, Python, and R providing the rules is available via a package registry and/or github: \href{https://github.com/NittanyLion/Quadriceps.jl}{Quadriceps.jl} (both 64-bit and 128-bit), \href{https://github.com/NittanyLion/quadriceps-py}{quadriceps-py} (64-bit only), and \href{https://github.com/NittanyLion/quadriceps-r}{quadriceps-r} (64-bit only).  The software used to create these rules is available via \href{https://github.com/NittanyLion/PositiveWeightQuadratureSolvers.jl}{PositiveWeightQuadratureSolvers.jl}.  The replication package is at \href{https://github.com/NittanyLion/QuadricepsReplicationPackage.jl}{QuadricepsReplicationPackage.jl}.  A snapshot of all five packages is archived at \url{https://doi.org/10.5281/zenodo.22883240}.}

\begin{document}

\title{Positive weight Hermite and Legendre quadrature rules}
\author{Joris Pinkse\thanks{\href{joris@psu.edu}{joris@psu.edu}}\\ 
Department of Economics\\ 
Penn State}
\date{September 2026}
\maketitle
\thispagestyle{empty}

\begin{abstract}
\headlineresults
\end{abstract}

\clearpage
\onehalfspacing

\section{Introduction}

\headlineresults\footnote{The Python and R packages are 64-bit only: neither language has a standard floating point type beyond double precision.  Julia has none either, but its generic typing lets Quadriceps.jl return a rule in whatever number type the caller asks for, \texttt{Float128} through Quadmath.jl or \texttt{BigFloat} from Base.}

The objective, for given $f,d$ and odd $p$, is to construct a rule of $N$ (node, weight) pairs $(n,w)$ for which
\begin{equation} \label{eq:integral equality}
∫ ∏_{j=1}ᵈ xⱼ^{aⱼ} f(x) \mathrm{d}x = ∑_{k=1}^N wₖ ∏_{j=1}ᵈ n_{kj}^{aⱼ}
\end{equation}
for every vector $a$ of integers with $aⱼ ≥ 0$ and $|a| = ∑_{j=1}ᵈ aⱼ ≤ p$.  These `rules' can then be used to approximate $∫ g(x) f(x) \mathrm{d}x$ by $∑_{k=1}^N wₖ g(nₖ)$.  The results in this paper apply to the case in which $f$ is one of the standard normal (Gauss-Hermite) and standard uniform (Gauss-Legendre) density functions.\footnote{The Gauss-Legendre domain is $[0,1]ᵈ$ in this paper.}   

I focus on the case of positive weights such that the integral approximation can be used as an argument to a logarithmic function.  In particular, this paper is motivated by random coefficients logit models as they are used in economics where the integrals are choice probabilities that enter a loglikelihood function.

Positive weights matter beyond ensuring that the approximation lies between the minimum and maximum function values of the function being integrated.  Positive weights ensure that there is no loss of digits due to cancellation of positive and negative weight contributions.  It has been shown \citep[see e.g.][section 4.1]{glaubitz21} that in a cube, the error of a quadrature rule is at most $1+∑ₖ|wₖ|$ times the error of the best uniform approximation by a polynomial of the same degree: only if the weights are nonnegative is $∑ₖ |wₖ| = ∑ₖ wₖ = 1$.  For the sparse grids that \citet{heiss08} provide for the cube, $∑ₖ |wₖ|$ is as much as 34 ($d=5$, $p=21$).

The easiest way of finding $(n,w)$ satisfying \cref{eq:integral equality} is to solve the problem in one dimension and then take the tensor product of the solution.  Doing so ensures positive weights, but it also requires $N = q^d$ nodes where $q=(p+1)/2$, i.e.\ the number of nodes then grows exponentially in $d$.  

Except in very low $p$ cases, there likely exist solutions to \cref{eq:integral equality} that require fewer nodes since the tensor product imposes more than just the restrictions in \cref{eq:integral equality}.  Indeed, if for instance $d=2,p=7$ then for the tensor product rule, \cref{eq:integral equality} also holds exactly when $a₁=a₂ = 6$, although $a₁+a₂ =12>7 = p$.  The number of nodes can be reduced significantly by letting go of cross moments of order greater than $p$.  The hardest part of this exercise is to achieve a low node count while imposing the positive-weight constraint.

None of the above is to say that the extra constraints provided by the tensor product have no value.  Indeed, one would expect $∫ g(x) f(x) \mathrm{d}x$ to be approximated better by a tensor product-based approximation than by a reduced node quadrature rule of the same order.  But there is no reason to believe that the tensor product provides the optimal approximation for a given node count.

I also make no claim that a plain Gauss-Hermite or Gauss-Legendre rule is necessarily the optimal way to do numerical integration.  Accuracy can often be improved by transforming the integrand before a rule is applied, for instance by recentering and rescaling it around its mode, as in adaptive Gauss-Hermite quadrature \citep{naylor82,liu94}, or by splitting a bounded  domain into subregions and applying a rule to each \citep{genz80}.  But such improvements are not the focus of this paper: they can be used in conjunction with the rules provided here, also.

I am not the first to study this problem.  Indeed, I will be making use of, combining, and building on methodology developed by others.  A complete description of the methodology used, including references, can be found in \cref{app:methods}.  The methodology used combines the following strategies.  It attempts to find a (nodes, weights) pair that is a solution to moment equations of (an orthogonalized version of) the form \cref{eq:integral equality}.  It eliminates nodes by exploiting symmetry, by trying to merge nodes, and by trying to find a new solution by reoptimizing after dropping nodes with small weights.  I further tried imposing symmetry: imposing symmetry lightens the computational load, but can increase the minimal achievable node count.  

I employed various warm-start strategies.  First, I used a laddering approach by starting for a fixed dimension $d$ from a lower value of $p$ and adding nodes from there.  I further tried starting from the tensor product of a $(d₁,p)$ and a $(d₂,p)$ solution to find a good $(d₁+d₂,p)$ solution. I tried taking a monotonic transformation from the Gauss-Hermite solution to obtain a starting point for the Gauss-Legendre solution and vice versa.  Finally, I used existing rules as a warm start and improved on them from there.

Any rules found were polished using a high-precision optimizer.

The computations were conducted over a six-week period on a combination of the Open Science Pool, Penn State's ROAR cluster, my office and home desktops, and my work and personal laptops.

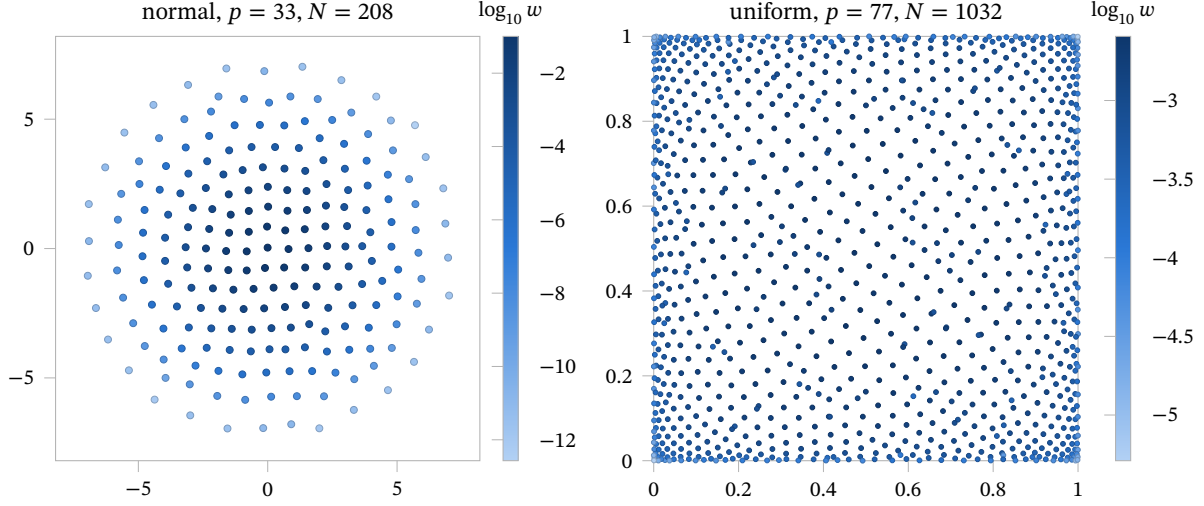
\begin{figure}[ht]
\centering
\begin{tikzpicture}
\begin{groupplot}[group style={group size=2 by 1, horizontal sep=2.3cm}, quad, scale only axis, width=0.34\linewidth, height=0.34\linewidth,
    grid=none, colormap={qblue}{color=(qd2!45) color=(qd3) color=(qd5)}, colorbar,
    colorbar style={width=2mm, tick label style={font=\scriptsize}, title={$\log_{10} w$}, title style={font=\scriptsize}, axis line style={gray!60}}]
  \nextgroupplot[title={normal, $p=33$, $N=208$}, xmin=-8.2, xmax=8.2, ymin=-8.2, ymax=8.2]
    \addplot[scatter, only marks, mark=*, mark size=1.3pt, scatter src=explicit, mark options={draw=gray!70, line width=0.15pt}] table[x=x, y=y, meta=logw] {fig/nodes_gh.dat};
  \nextgroupplot[title={uniform, $p=77$, $N=1032$}, xmin=0, xmax=1, ymin=0, ymax=1]
    \addplot[scatter, only marks, mark=*, mark size=0.9pt, scatter src=explicit, mark options={draw=gray!70, line width=0.1pt}] table[x=x, y=y, meta=logw] {fig/nodes_le.dat};
\end{groupplot}
\end{tikzpicture}
\caption{The largest planar rule of each table: nodes, shaded by the logarithm of their weight.}
\label{fig:nodes}
\end{figure}

As will become clear in \cref{sec:results}, the tables in this paper improve on the rules available in the literature for many cases and provide new ones for many cases for which no published rules (other than the tensor product) exist.  For Gauss-Hermite, I develop entirely new or improved rules in 34 of 57 cases whereas for Gauss-Legendre the count is 62 of 85.

Although my primary objective was to improve the existing Gauss-Hermite results, my Gauss-Legendre rules cover more cases.  This is simply due to the fact that finding a Gauss-Hermite solution is harder because of the presence of (node, weight) combinations that are extreme in their location (nodes) and magnitude (weight).  This difference in the spread of the weights between Gauss-Hermite and Gauss-Legendre solutions is illustrated in \cref{fig:nodes}.

\section{Results}
\label{sec:results}

\begin{table}[p]
\centering
\setlength{\tabcolsep}{3.5pt}
\small
\begin{tabular}[t]{rrrrrrrc}
\toprule
$p$ & $N$ & $\rho$ & rel.\ err. & M\"o & sym & prev & src \\
\midrule
\multicolumn{8}{@{}l}{$d = 2$} \\
\rowcolor{gray!40}1 & 1 & 1.00 & 0 & 1 & 1 & 1 & --- \\
\rowcolor{gray!40}3 & 4 & 1.00 & 3.6e-70 & 4 & 4 & 4 & CO \\
\rowcolor{gray!40}5 & 7 & 0.88 & 2.7e-69 & 7 & 7 & 7 & CO \\
\rowcolor{gray!40}7 & 12 & 0.87 & 1.2e-69 & 12 & 12 & 12 & CO \\
\rowcolor{gray!15}9 & 18 & 0.85 & 6.8e-69 & 17 & 17 & 18 & HP \\
\rowcolor{gray!15}11 & 25 & 0.83 & 1.1e-79 & 24 & 25 & 25 & HP \\
\rowcolor{gray!15}13 & 34 & 0.83 & 8.9e-80 & 31 & 33 & 34 & CH \\
\rowcolor{gray!15}15 & 44 & 0.83 & 1.3e-79 & 40 & 44 & 44 & CO \\
17 & 55 & 0.82 & 6.4e-79 & 49 & 55 & 57 & SC \\
19 & 68 & 0.82 & 6.8e-78 & 60 & 68 & 71 & CO \\
21 & 82 & 0.82 & 3.0e-77 & 71 & 81 & 90 & CO \\
\rowcolor{gray!15}23 & 97 & 0.82 & 8.7e-74 & 84 & 97 & 97 & CO \\
25 & 114 & 0.82 & 6.5e-71 & 97 & 113 & 127 & CO \\
27 & 132 & 0.82 & 1.1e-69 & 112 & 132 & --- & --- \\
29 & 153 & 0.82 & 5.0e-78 & 127 & 151 & --- & --- \\
31 & 178 & 0.83 & 7.9e-69 & 144 & 172 & --- & --- \\
33 & 208 & 0.85 & 7.2e-69 & 161 & 193 & --- & --- \\
\midrule
\multicolumn{8}{@{}l}{$d = 3$} \\
\rowcolor{gray!40}1 & 1 & 1.00 & 0 & 1 & 1 & 1 & --- \\
\rowcolor{gray!40}3 & 6 & 0.91 & 5.4e-71 & 6 & 6 & 6 & CO \\
\rowcolor{gray!40}5 & 13 & 0.78 & 1.1e-69 & 13 & 13 & 13 & ST \\
\rowcolor{gray!15}7 & 27 & 0.75 & 6.0e-69 & 26 & 27 & 27 & CO \\
\rowcolor{gray!15}9 & 45 & 0.71 & 9.3e-69 & 43 & 45 & 45 & KO \\
\rowcolor{gray!15}11 & 77 & 0.71 & 1.0e-79 & 68 & 77 & 77 & CO \\
13 & 128 & 0.72 & 6.3e-80 & 99 & 119 & 137 & CO \\
15 & 184 & 0.71 & 5.2e-79 & 140 & 175 & --- & --- \\
17 & 264 & 0.71 & 2.6e-78 & 189 & 248 & --- & --- \\
19 & 354 & 0.71 & 1.4e-69 & 250 & 342 & --- & --- \\
21 & 476 & 0.71 & 1.8e-71 & 321 & 447 & --- & --- \\
23 & 597 & 0.70 & 3.0e-72 & 406 & 563 & --- & --- \\
25 & 776 & 0.71 & 5.1e-73 & 503 & 720 & --- & --- \\
27 & 966 & 0.71 & 2.6e-71 & 616 & 894 & --- & --- \\
29 & 1242 & 0.72 & 6.3e-72 & 743 & 1094 & --- & --- \\
31 & 1848 & 0.77 & 2.1e-71 & 888 & 1328 & --- & --- \\
33 & 2226 & 0.77 & 1.1e-69 & 1049 & 1588 & --- & --- \\
\bottomrule
\end{tabular}
\hfill
\begin{tabular}[t]{rrrrrrrc}
\toprule
$p$ & $N$ & $\rho$ & rel.\ err. & M\"o & sym & prev & src \\
\midrule
\multicolumn{8}{@{}l}{$d = 4$} \\
\rowcolor{gray!40}1 & 1 & 1.00 & 0 & 1 & 1 & 1 & --- \\
\rowcolor{gray!40}3 & 8 & 0.84 & 6.6e-70 & 8 & 8 & 8 & CO \\
\rowcolor{gray!15}5 & 22 & 0.72 & 1.8e-69 & 21 & 22 & 22 & CO \\
\rowcolor{gray!15}7 & 49 & 0.66 & 7.2e-69 & 48 & 49 & 49 & CO \\
9 & 116 & 0.66 & 7.7e-80 & 91 & 105 & --- & --- \\
11 & 193 & 0.62 & 8.3e-80 & 160 & 192 & --- & --- \\
13 & 414 & 0.64 & 6.7e-80 & 259 & 353 & --- & --- \\
15 & 577 & 0.61 & 7.2e-80 & 400 & 529 & --- & --- \\
17 & 1056 & 0.63 & 1.1e-77 & 589 & 856 & --- & --- \\
19 & 1505 & 0.62 & 1.5e-79 & 840 & 1281 & --- & --- \\
21 & 2318 & 0.63 & 1.9e-69 & 1161 & 1809 & --- & --- \\
23 & 3238 & 0.63 & 1.3e-71 & 1568 & 2512 & --- & --- \\
\midrule
\multicolumn{8}{@{}l}{$d = 5$} \\
\rowcolor{gray!40}1 & 1 & 1.00 & 0 & 1 & 1 & 1 & --- \\
\rowcolor{gray!40}3 & 10 & 0.79 & 2.2e-70 & 10 & 10 & 10 & CO \\
\rowcolor{gray!15}5 & 32 & 0.67 & 2.1e-69 & 31 & 32 & 32 & SS \\
\rowcolor{gray!15}7 & 83 & 0.61 & 8.2e-80 & 80 & 83 & 83 & ST \\
9 & 244 & 0.60 & 5.4e-80 & 171 & 212 & 395 & AD \\
11 & 485 & 0.57 & 9.7e-80 & 332 & 445 & --- & --- \\
13 & 1135 & 0.58 & 1.9e-79 & 591 & 846 & --- & --- \\
15 & 1767 & 0.56 & 5.5e-80 & 992 & 1496 & --- & --- \\
17 & 3986 & 0.58 & 1.4e-79 & 1581 & 2448 & --- & --- \\
19 & 7174 & 0.59 & 3.0e-79 & 2422 & 3978 & --- & --- \\
21 & 13199 & 0.61 & 9.8e-80 & 3583 & 6290 & --- & --- \\
\bottomrule
\end{tabular}
\normalsize
\caption{Positive-weight rules for the Gaussian weight $N(0, I_d)$. $N$: nodes in my best rule; $\rho = N^{1/d}/q$ with $q = (p+1)/2$ ($\rho = 1$ is the Gauss product grid, smaller is better); rel.\ err.: largest relative monomial error, $\max_{|a| \le p} |\sum_s w_s x_s^a - \mathrm{E}\,x^a| / \max(\sum_s w_s\,|x_s^a|, 1)$, of the rule's 80-digit file (rounded to double precision: at most $1.1\cdot 10^{-15}$); M\"o: M\"oller's lower bound; sym: the symmetric counting floor, the fewest nodes for which a rule invariant under a group $G$ has as many unknowns as it has $G$-invariant moment conditions, overall and among the conditions that only some orbit types see, minimized over the orbit types and over $G$ (no symmetry, $\pm$pairs, a sign change in each coordinate, the symmetry group $B_d$ of the cube and, for $d = 2$, the dihedral and cyclic groups), not counting the $d(d-1)/2$ unknowns that only rotate the rule where $G$ commutes with rotations --- a parameter count, not a proven bound; prev: best count in the literature, --- where none exists besides the product grid; src: where that count is recorded --- the original paper where I could identify it, otherwise a compilation (AD = \citealt{adurthi12}; CH = \citealt{coolio88b}; CO = \citealt{coolio}; HP = \citealt{haggy76,haggy77}; KO = \citealt{konyaev77}; SC = \citealt{degani05}; SS = \citealt{stroud63}; ST = \citealt{stroud71}). Light gray: no improvement on prev; dark gray: M\"oller's bound attained (proven minimal).}\label{tab:best-gh}
\end{table}

\begin{table}[p]
\centering
\setlength{\tabcolsep}{3.5pt}
\footnotesize
\begin{tabular}[t]{rrrrrrrc}
\toprule
$p$ & $N$ & $\rho$ & rel.\ err. & M\"o & pair & prev & src \\
\midrule
\multicolumn{8}{@{}l}{$d = 2$} \\
\rowcolor{gray!40}1 & 1 & 1.00 & 0 & 1 & 1 & 1 & --- \\
\rowcolor{gray!40}3 & 4 & 1.00 & 5.2e-75 & 4 & 3 & 4 & ST \\
\rowcolor{gray!40}5 & 7 & 0.88 & 1.4e-76 & 7 & 6 & 7 & ST \\
\rowcolor{gray!40}7 & 12 & 0.87 & 1.1e-74 & 12 & 11 & 12 & ST \\
\rowcolor{gray!40}9 & 17 & 0.82 & 1.4e-78 & 17 & 17 & 17 & MO \\
\rowcolor{gray!40}11 & 24 & 0.82 & 4.7e-73 & 24 & 24 & 24 & CH \\
\rowcolor{gray!15}13 & 33 & 0.82 & 4.1e-72 & 31 & 33 & 33 & CH \\
\rowcolor{gray!15}15 & 43 & 0.82 & 6.7e-74 & 40 & 43 & 43 & FS \\
\rowcolor{gray!15}17 & 54 & 0.82 & 1.5e-78 & 49 & 54 & 54 & FS \\
\rowcolor{gray!15}19 & 67 & 0.82 & 1.4e-71 & 60 & 67 & 67 & FS \\
\rowcolor{gray!15}21 & 81 & 0.82 & 1.4e-78 & 71 & 81 & 81 & FS \\
\rowcolor{gray!15}23 & 96 & 0.82 & 2.5e-78 & 84 & 96 & 96 & FS \\
\rowcolor{gray!15}25 & 113 & 0.82 & 1.9e-70 & 97 & 113 & 113 & FS \\
27 & 131 & 0.82 & 1.2e-70 & 112 & 131 & 132 & FS \\
29 & 150 & 0.82 & 1.3e-78 & 127 & 150 & 152 & XG \\
31 & 171 & 0.82 & 2.7e-72 & 144 & 171 & 172 & FS \\
33 & 194 & 0.82 & 2.7e-73 & 161 & 193 & 197 & FS \\
35 & 216 & 0.82 & 1.9e-77 & 180 & 216 & 220 & FS \\
37 & 242 & 0.82 & 5.3e-76 & 199 & 241 & 245 & FS \\
39 & 268 & 0.82 & 3.1e-73 & 220 & 267 & 274 & FS \\
41 & 294 & 0.82 & 3.8e-74 & 241 & 294 & 303 & FS \\
43 & 324 & 0.82 & 8.6e-73 & 264 & 323 & 331 & FS \\
45 & 356 & 0.82 & 8.9e-72 & 287 & 353 & 359 & FS \\
47 & 388 & 0.82 & 2.1e-71 & 312 & 384 & 396 & FS \\
49 & 422 & 0.82 & 2.5e-78 & 337 & 417 & 427 & FS \\
51 & 456 & 0.82 & 8.1e-73 & 364 & 451 & 462 & FS \\
53 & 492 & 0.82 & 7.1e-75 & 391 & 486 & 498 & FS \\
55 & 528 & 0.82 & 2.7e-70 & 420 & 523 & 536 & FS \\
57 & 570 & 0.82 & 2.0e-71 & 449 & 561 & 576 & DW \\
59 & 606 & 0.82 & 1.2e-72 & 480 & 600 & 613 & DW \\
61 & 642 & 0.82 & 2.0e-78 & 511 & 641 & 660 & DW \\
63 & 692 & 0.82 & 1.1e-74 & 544 & 683 & 709 & DW \\
65 & 732 & 0.82 & 5.7e-71 & 577 & 726 & 757 & DW \\
67 & 780 & 0.82 & 2.0e-73 & 612 & 771 & 805 & DW \\
69 & 826 & 0.82 & 1.0e-70 & 647 & 817 & 849 & DW \\
71 & 872 & 0.82 & 2.1e-73 & 684 & 864 & 904 & DW \\
73 & 922 & 0.82 & 1.5e-70 & 721 & 913 & 953 & DW \\
75 & 980 & 0.82 & 2.1e-70 & 760 & 963 & 1001 & DW \\
77 & 1032 & 0.82 & 6.9e-74 & 799 & 1014 & 1049 & DW* \\
\midrule
\multicolumn{8}{@{}l}{$d = 3$} \\
\rowcolor{gray!40}1 & 1 & 1.00 & 0 & 1 & 1 & 1 & --- \\
\rowcolor{gray!40}3 & 6 & 0.91 & 1.1e-76 & 6 & 4 & 6 & ST \\
\rowcolor{gray!40}5 & 13 & 0.78 & 4.9e-76 & 13 & 12 & 13 & ST \\
\rowcolor{gray!40}7 & 26 & 0.74 & 6.9e-76 & 26 & 26 & 26 & XG \\
\bottomrule
\end{tabular}
\hfill
\begin{tabular}[t]{rrrrrrrc}
\toprule
$p$ & $N$ & $\rho$ & rel.\ err. & M\"o & pair & prev & src \\
\midrule
\multicolumn{8}{@{}l}{$d = 3$ (cont.)} \\
9 & 48 & 0.73 & 6.9e-75 & 43 & 48 & 50 & XG \\
11 & 82 & 0.72 & 1.9e-74 & 68 & 81 & 84 & XG \\
13 & 128 & 0.72 & 5.7e-74 & 99 & 126 & 130 & XG \\
15 & 188 & 0.72 & 6.9e-74 & 140 & 186 & 190 & XG \\
17 & 266 & 0.71 & 4.2e-73 & 189 & 263 & 282 & DW \\
19 & 360 & 0.71 & 9.6e-73 & 250 & 358 & 369 & DW \\
21 & 476 & 0.71 & 1.5e-71 & 321 & 474 & 495 & DW \\
23 & 612 & 0.71 & 7.2e-70 & 406 & 612 & 617 & DW \\
25 & 776 & 0.71 & 2.2e-71 & 503 & 774 & 828 & DW \\
27 & 964 & 0.71 & 2.2e-75 & 616 & 963 & 984 & DW* \\
29 & 1184 & 0.71 & 9.3e-78 & 743 & 1180 & 1258 & DW* \\
31 & 1432 & 0.70 & 2.8e-76 & 888 & 1428 & 1478 & DW* \\
33 & 1714 & 0.70 & 6.7e-78 & 1049 & 1709 & 1787 & DW* \\
35 & 2028 & 0.70 & 2.6e-76 & 1230 & 2024 & 2102 & DW* \\
37 & 2380 & 0.70 & 1.0e-75 & 1429 & 2376 & 2506 & DW* \\
39 & 2770 & 0.70 & 2.7e-74 & 1650 & 2766 & 2856 & DW* \\
41 & 3200 & 0.70 & 7.9e-75 & 1891 & 3196 & 3338 & DW* \\
43 & 3704 & 0.70 & 4.9e-71 & 2156 & 3669 & 3870 & DW* \\
45 & 4308 & 0.71 & 2.3e-71 & 2443 & 4186 & 4414 & DW* \\
\midrule
\multicolumn{8}{@{}l}{$d = 4$} \\
\rowcolor{gray!40}1 & 1 & 1.00 & 0 & 1 & 1 & 1 & --- \\
\rowcolor{gray!40}3 & 8 & 0.84 & 2.2e-77 & 8 & 5 & 8 & ST \\
\rowcolor{gray!40}5 & 21 & 0.71 & 4.7e-76 & 21 & 19 & 21 & KK \\
7 & 54 & 0.68 & 2.4e-75 & 48 & 52 & 55 & KK \\
9 & 120 & 0.66 & 2.0e-74 & 91 & 118 & 138 & KK \\
11 & 234 & 0.65 & 5.4e-73 & 160 & 233 & 272 & FR \\
13 & 416 & 0.65 & 5.9e-72 & 259 & 415 & 512 & FR \\
15 & 690 & 0.64 & 5.9e-71 & 400 & 687 & 728 & FR \\
17 & 1078 & 0.64 & 3.9e-71 & 589 & 1074 & 1384 & FR \\
19 & 1612 & 0.63 & 2.7e-70 & 840 & 1606 & --- & --- \\
21 & 2322 & 0.63 & 5.9e-71 & 1161 & 2315 & --- & --- \\
23 & 3244 & 0.63 & 8.0e-76 & 1568 & 3235 & --- & --- \\
\midrule
\multicolumn{8}{@{}l}{$d = 5$} \\
\rowcolor{gray!40}1 & 1 & 1.00 & 0 & 1 & 1 & 1 & --- \\
\rowcolor{gray!40}3 & 10 & 0.79 & 4.1e-77 & 10 & 6 & 10 & ST \\
\rowcolor{gray!15}5 & 32 & 0.67 & 1.2e-75 & 31 & 30 & 32 & ST \\
7 & 100 & 0.63 & 2.9e-73 & 80 & 100 & --- & --- \\
9 & 266 & 0.61 & 4.4e-72 & 171 & 264 & --- & --- \\
11 & 602 & 0.60 & 2.2e-72 & 332 & 598 & --- & --- \\
13 & 1212 & 0.59 & 4.6e-73 & 591 & 1204 & --- & --- \\
15 & 2500 & 0.60 & 4.6e-73 & 992 & 2224 & --- & --- \\
17 & 3872 & 0.58 & 1.2e-71 & 1581 & 3840 & --- & --- \\
19 & 6826 & 0.58 & 1.4e-78 & 2422 & 6278 & --- & --- \\
21 & 10984 & 0.58 & 1.1e-78 & 3583 & 9820 & --- & --- \\
\bottomrule
\end{tabular}
\normalsize
\caption{Positive-weight rules for the uniform weight on $[0,1]^d$. $N$: nodes in my best rule; $\rho = N^{1/d}/q$ with $q = (p+1)/2$ ($\rho = 1$ is the Gauss product grid, smaller is better); rel.\ err.: largest relative monomial error, $\max_{|a| \le p} |\sum_s w_s x_s^a - \mathrm{E}\,x^a| / \max(\sum_s w_s\,|x_s^a|, 1)$, of the rule's 80-digit file (rounded to double precision: at most $3.7\cdot 10^{-17}$); M\"o: M\"oller's lower bound; pair: the counting floor for a $\pm$pair rule, $2\lceil M_{\mathrm{even}}/(d+1) \rceil$ with $M_{\mathrm{even}}$ the number of monomials of even total degree $\le p$ (one less where a node at the origin allows it) --- a parameter count for that ansatz, not a proven bound; prev: best count in the literature, --- where none exists besides the product grid; src: where that count is recorded --- the original paper where I could identify it, otherwise a compilation (CH = \citealt{coolio88b}; DW = \citealt{diablo}; FR = \citealt{frontin21}; FS = \citealt{fiesta}; KK = \citealt{cash}; MO = \citealt{molly76}; ST = \citealt{stroud71}; XG = \citealt{xiao10}); *: my rule was warm-started from theirs. Light gray: no improvement on prev; dark gray: M\"oller's bound attained (proven minimal). Every such file also meets the stopping criterion of \citealt{diablo}, $\|f - V^{\top} w\|_2 < 10^{-66}$, over the full Legendre basis (appendix).}\label{tab:best-le}
\end{table}

A description of the results for the Gauss-Hermite and Gauss-Legendre cases can be found in \cref{tab:best-gh,tab:best-le}, respectively.  Each table segment corresponds to a different dimension $d$.  Each row in a table segment corresponds to a different polynomial exactness degree $p$.  

There are four columns with node counts in each table: $N$ is the number of nodes used by the proposed rule, Mö is Möller's lower bound \citep{molly79}, and prev refers to the node count of the best available rule prior to this paper.  The Möller bound is attainable for some combinations of $d,p$ but for most cases with $d≥3$ it is unknown and doubtful that rules achieving the Möller bound exist.  So, the fact that a rule does not attain the Möller lower bound does not imply that a better rule exists.   The fourth node count column in each table (pair in \cref{tab:best-le} and sym in \cref{tab:best-gh}) is a counting floor rather than a bound: \cref{app:methods} sets out how each floor is counted, and what it does and does not imply.  Again, these floors are not known lower bounds (and in some cases are even less than the Möller bounds), but are a better reflection of what is achievable than the Möller bounds if $d$ or $p$ is large.

The constraints imposed are not the same across papers.  For instance, \citet[\citetalias{diablo}]{diablo} impose a greater degree of symmetry on the solution than some of the methods I use do, so it is not surprising that the node counts that I provide are lower than those in \citetalias{diablo}, except where \citetalias{diablo} attain the Möller bound. As mentioned in \cref{app:methods}, the rules provided by \citetalias{diablo} were moreover used to warm-start my algorithms. 

In addition, some papers contain some positive-weight rules even though weights were not restricted to be positive.  This is e.g.\ true for \citet{sandy} who, in contrast to most papers cited here, puts emphasis on even $p$ cases.  None of the \citet{sandy} rules provide the lowest node counts for the (odd) $p$ values presented here.

Another column in the tables contains the value of $ρ$, which reflects the equivalent one-dimensional count in the sense that the number $ρ=0.60$ in \cref{tab:best-gh} for $d=5$, $p=9$ ($q=5$) means that the node count in that case is equivalent to that of the tensor product for $q=3$ ($3^5 = 243 ≈ 244$).  
\begin{figure}[t]
\centering
\begin{tikzpicture}
\begin{groupplot}[group style={group size=2 by 1, horizontal sep=1.2cm}, quad, width=0.5\linewidth, height=0.4\linewidth,
    ymin=0.5, ymax=0.92, xmin=5, xlabel={degree $p$}, restrict x to domain=5:99]
  \nextgroupplot[title={normal}, ylabel={$\rho$}, ylabel style={rotate=-90, font=\normalsize}, xmax=38.5]
    \addplot[qd2, line width=0.9pt, mark=*, mark size=1pt] table[x=p, y=rho] {fig/rho_gh_d2.dat};
    \addplot[qd2, line width=0.6pt, densely dashed] table[x=p, y=floor] {fig/rho_gh_d2.dat};
    \pgfplotsinvokeforeach{3,4,5}{
      \addplot[qd#1, line width=0.9pt, mark=*, mark size=1pt, forget plot] table[x=p, y=rho] {fig/rho_gh_d#1.dat};
      \addplot[qd#1, line width=0.6pt, densely dashed, forget plot] table[x=p, y=floor] {fig/rho_gh_d#1.dat};
    }
    \node[anchor=west, font=\scriptsize] at (axis cs:33.4,0.852) {$d=2$};
    \node[anchor=west, font=\scriptsize] at (axis cs:33.4,0.765) {$d=3$};
    \node[anchor=west, font=\scriptsize] at (axis cs:23.4,0.632) {$d=4$};
    \node[anchor=west, font=\scriptsize] at (axis cs:21.4,0.598) {$d=5$};
  \nextgroupplot[title={uniform}, xmax=93]
    \pgfplotsinvokeforeach{2,3,4,5}{
      \addplot[qd#1, line width=0.9pt, mark=*, mark size=0.8pt] table[x=p, y=rho] {fig/rho_le_d#1.dat};
      \addplot[qd#1, line width=0.6pt, densely dashed] table[x=p, y=floor] {fig/rho_le_d#1.dat};
    }
    \node[anchor=west, font=\scriptsize] at (axis cs:77.5,0.824) {$d=2$};
    \node[anchor=west, font=\scriptsize] at (axis cs:45.5,0.712) {$d=3$};
    \node[anchor=west, font=\scriptsize] at (axis cs:23.5,0.632) {$d=4$};
    \node[anchor=west, font=\scriptsize] at (axis cs:21.5,0.578) {$d=5$};
\end{groupplot}
\end{tikzpicture}
\caption{The node count of each rule for $p≥5$ in \cref{tab:best-gh,tab:best-le} as a fraction of the Gauss product grid's, per axis: $\rho = N^{1/d}/q$ with $q = (p+1)/2$ (solid, one marker per rule), so that $\rho = 1$ is the product grid, and the counting floor of the tables' sixth column on the same scale (dashed).}
\label{fig:rho}
\end{figure}
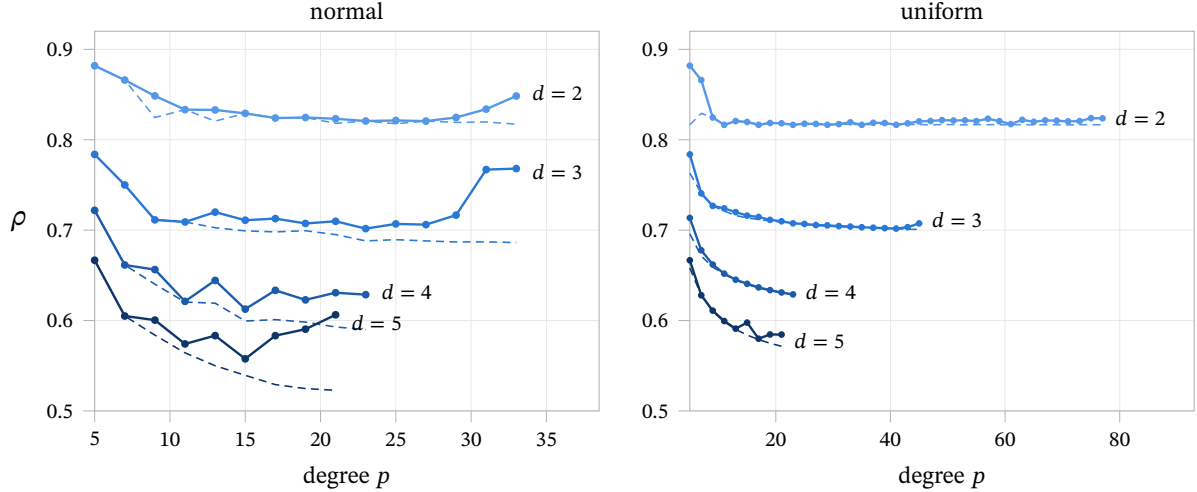
As \cref{fig:rho} shows, in each table the $ρ$ values for a fixed value of $d$ flatten out as $p$ increases.  The fact that the $ρ$ values tick up for larger values of $p$ in \cref{tab:best-gh} may suggest that those rules are suboptimal. The value of $p$ at which I stopped producing rules for a given value of $d$ is thus chosen as the value of $p≥21$ at which the value of $ρ$ started to edge up.

Each table contains two more columns: one displaying the relative error of the result and one the source where I found the previous best result.\footnote{Cools' Encyclopaedia of Cubature Formulas is the recorded source for sixteen of the Gauss-Hermite counts in \cref{tab:best-gh}.  Its overview tables are online; I requested access to the full collection and received no reply, so those counts are as the Encyclopaedia records them rather than checked against the original publications.}

Dark cell coloring means that rules attaining the Möller lower bound already exist and light cell coloring that I have not found rules improving on what exists, albeit that my solution can differ from the preexisting results.

That said, most of the entries have white backgrounds, reflecting the case in which my results either improve on what exists or are the first to provide a solution different from the tensor product at all.  Here, as throughout the paper, only positive-weight rules count: sparse-grid rules \citep{smolyak63}, for the normal density typically built on the nested one-dimensional rules of \citet{genz96}, exist for every $d$ and $p$, but they generally have some negative weights.

\clearpage
\appendix
\crefalias{section}{appendix}

\providecommand{\origin}{\textit{Origin:}~}
\providecommand{\used}{\textit{Used here:}~}

\section{Methods used for the Gaussian and uniform-weight rule tables, with attribution}
\label{app:methods}

\noindent
This list covers every method that contributed to, or was tried for, the positive-weight
rules for the Gaussian weight $N(0,I_d)$ (GH) and the uniform weight on $[0,1]^d$ (Le). For
each method it says what the method does, whom it comes from, and what it was used for
here. ``This project'' marks a method, or a variant, for which I found no earlier source.
That is a statement about my literature search, not a priority claim.

\subsection{Solving the moment equations}

\begin{description}[style=nextline,leftmargin=1.5em,itemsep=0.6ex]
\item[Designed-quadrature formulation]
  All node coordinates and weights are unknowns, and the residual is the moment error in an
  orthonormal polynomial basis, with penalty rows for positivity (and, on the cube, for the
  box).  The first solver was a Julia translation of the Matlab code accompanying the paper.
  \origin \citet{cash}.
  \used the starting point for every solver below. I use orthonormalized Hermite and
  Legendre bases, because raw monomials wreck the conditioning at high degree.

\item[Levenberg-Marquardt with gain-ratio damping]
  Damped Gauss-Newton on the moment residual, with Nielsen's update of the damping
  parameter.
  \origin \citet{levenberg44}, \citet{marquardt63}; damping update \citet{nielsen99},
  \citet{madsen04}.
  \used the inner solve of every search.

\item[Geodesic acceleration]
  A second-order correction along the LM step, from one extra residual evaluation.
  \origin \citet{transtrum12}.
  \used in the free-node and pair solvers.

\item[Log-weight parameterization]
  Weights are written as $w = e^{u}$, so positivity holds by construction and the penalty
  rows and barrier parameter go away.
  \origin a standard reparameterization; \citetalias{diablo} likewise enforce positivity (and
  interiority) through the choice of variables in a Levenberg-Marquardt solver.
  \used all free-node and pair solves.

\item[Multithreaded pair elimination]
  The pair solver with multithreaded linear algebra, normal equations formed by a
  symmetric rank-$k$ update, checkpoints inside a solve, and Newton steps through the dual
  Cholesky factor for large rules.
  \origin this project (engineering of the solver above).
  \used Le $d=4$, $p=21, 23$ and $d=5$, $p=17$--$21$. Before this, the single-threaded solver
  needed about 20 hours for one solve at these sizes and never finished one.
\end{description}

\subsection{Removing nodes}

\begin{description}[style=nextline,leftmargin=1.5em,itemsep=0.6ex]
\item[Node elimination]
  Start from an exact rule, drop the lightest node, reconverge, and repeat until no drop
  reconverges.
  \origin \citet{xiao10} (with a significance-ordered candidate list); point elimination on
  the square, \citet{fiesta}; node elimination for polytopes, \citet{slobodkins23}.
  \used most GH and Le rules at $d=2$ and the free-node searches at $d=3$. The free-node
  eliminator is, apart from the choice of starting set, the same method as Festa and
  Sommariva's.

\item[Pair (centrally symmetric) elimination]
  Nodes are constrained to $\pm$ pairs about the center, which satisfies every odd moment
  identically and halves the unknowns; the eliminator then drops one pair at a time.
  \origin central symmetry as an ansatz is classical \citep{stroud71}; orbit elimination
  and orbit collapse for symmetric rules, \citetalias{diablo}.
  \used the workhorse for Le at every dimension, and for GH at $d=2$.

\item[Merge move]
  Drive two nodes into coincidence, which removes several nodes at once and reaches basins
  that no sequence of single drops reaches.
  \origin inspired by \citet{haggy76}, who reduced a degree-11 Gaussian rule from 28 to 25
  nodes by letting orbits coincide at the origin; its use as a step in node elimination is
  this project's.
  \used inside free-node elimination.

\item[Center-node merge]
  A pairs-only rule has an even count. Where the counting bound with a center node is one
  below the pairs-only bound, collapse the innermost pair into a center node and
  reconverge.
  \origin this project.
  \used Le $d=2$, $p=19, 21, 27, 31$ (one node each).
\end{description}

\subsection{Imposed symmetry}

\begin{description}[style=nextline,leftmargin=1.5em,itemsep=0.6ex]
\item[Invariant-moment reduction]
  A rule invariant under a finite group is exact to degree $p$ if and only if it is exact
  on the invariant polynomials of degree at most $p$. Orbits through special positions
  carry many nodes for few parameters.
  \origin \citet{sobolev62}; fully symmetric rules, \citet{stroud71}.
  \used the basis of every orbit search below.

\item[$B_d$ orbit search]
  Signed permutations of the coordinates, with orbit-constant weights. The search runs over
  multisets of orbit types, then degenerates the result by dropping an orbit, collapsing a
  value group onto a coordinate subspace, or merging two values.
  \origin the ansatz is Sobolev's; the degeneration moves are this project's, and the
  orbit collapse is close to \citetalias{diablo}.
  \used the broad GH and Le front at $d=3$--$5$ (for example GH $d=5$, $p=11$--$21$).

\item[$D_d$ half-orbits, $S_d \times \mathbb{Z}_2$, simplex $S_{d+1}\times\mathbb{Z}_2$]
  Symmetry classes that $B_d$ cannot express: even sign changes only; permutations with
  central inversion; the symmetry group of the simplex.
  \origin these symmetry classes appear throughout \citet{stroud71}.
  \used $S_d\times\mathbb{Z}_2$ produced GH $d=4$, $p=9$ ($N=116$) and the $d=5$, $p=9$
  chain ($N=244$). The simplex ansatz re-derived Stroud and Secrest's 32-node rule
  \citep{stroud63} cold.

\item[Exceptional groups]
  Icosahedral symmetry at $d=3$; the 24-cell group $F_4$ and the icosian group $H_4$ at
  $d=4$.
  \origin icosahedral rules for the Gaussian weight, \citet{stroud71} and \citet{konyaev77}.
  \used as validation: the search recovered Stroud's 13-node rule and Konyaev's 45-node
  rule (GH $d=3$, $p=9$). That 45-node rule is Konyaev's, not mine. The $H_4$ search was
  retired after about 331{,}000 restarts found nothing feasible.

\item[Rotational $D_k$ orbits in the plane]
  For $N(0,I_2)$ every dihedral group $D_k$ is an admissible symmetry. Exactness is imposed
  on the $D_k$-invariant Laguerre-Fourier basis.
  \origin the reduction is Sobolev's; the best classical planar Gaussian rules have such
  symmetry, for instance the hexagonal 25-node degree-11 rule of \citet{haggy76}; the $D_k$
  search is this project's.
  \used searches at GH $d=2$, $p=31$--$35$; it has not produced a banked rule.

\item[Point-group enumeration]
  The finite subgroups of $O(3)$ (the seven polyhedral groups and the seven axial
  families, the latter up to axial order 12) and every composition of $N$ nodes into their
  orbits, each solved from random starts, with Molien's series counting the invariant
  equations for pruning. The cap loses nothing here: for $p=9$, a ring of 10 or more nodes
  already integrates every trigonometric polynomial that degree requires, so a solution
  with larger rings implies one with 10-node rings and fewer than 43 nodes, which Möller's
  bound excludes.
  \origin \citet{molien97} for the counting; the enumeration is this project's.
  \used GH $d=3$, $p=9$ at $N = 43, 44$, below Konyaev's 45: all 41{,}353 structured cases
  (roughly 3.5 passes of about 850 random starts each) and a symmetry-free catch-all
  (13{,}000--17{,}000 starts per case) found no rule; the best residual was about
  $8\cdot 10^{-3}$. This supports, but does not prove, that 45 is minimal.

\item[Enumeration certification of $B_d$ floors]
  An exhaustive, deterministic enumeration of $B_d$ orbit structures that certifies the
  smallest $B_d$-symmetric rule for a cell.
  \origin this project.
  \used $B_d$ floors for $d=3$, $p\le 15$ and $d=4$, $p\le 13$. These bound only the
  symmetric class, not all rules.
\end{description}

\subsection{Starting points}

\begin{description}[style=nextline,leftmargin=1.5em,itemsep=0.6ex]
\item[Gauss product grid]
  The tensor product of one-dimensional Gauss rules, computed by the Golub-Welsch
  eigenvalue method: exact, with $q^d$ nodes.
  \origin \citet{golub69}.
  \used a start of last resort, and the reference point $\rho = 1$ in the tables.

\item[Spectral initialization]
  Candidate nodes from the eigenstructure of truncated multiplication operators.
  \origin \citet{vioreanu14}; \citet{bremer10}.
  \used GH and Le starts. At Le $d=2$, $p=13$ it reached 33, the published count, where
  elimination from the published rule had stalled; the rule turned out to be Festa and
  Sommariva's own.

\item[Degree continuation]
  Warm-start degree $p$ from the degree $p-2$ rule, with rescaled radii.
  \origin \citet{xiao10}.
  \used GH $d=3$, $p=31$ and $33$ (and the Le degree ladders).

\item[Warm starts from published rules]
  Start elimination from somebody else's rule for the same cell.
  \origin the starting rules are those of \citetalias{diablo} and \citet{fiesta}.
  \used Le $d=2$, $p=77$ and $d=3$, $p=27$--$45$ start from the rules of \citetalias{diablo}:
  the counts in those rows are mine, but the starting points are theirs and are hereby
  credited. Warm starts from the rules of \citet{fiesta} (Le $d=2$, $p=13$--$25$) found
  nothing below their counts, so no tabulated rule descends from them. Four Le $d=2$ rules
  are theirs outright: $p=25$ ($N=113$) is transcribed, and $p=9$, $13$, $15$ ($N=17$, $33$,
  $43$) were re-derived by my searches and turned out to be theirs (the counts at $p=9$ and
  $13$ were first published by \citet{molly76} and \citet{coolio88b}, which is why the
  tables credit those two cells to them).

\item[Tchakaloff compression]
  Non-negative least squares on the moment system of a fine tensor rule, which returns a
  positive rule with at most as many nodes as moments.
  \origin \citet{tchakaloff57}; NNLS, \citet{lawson74}.
  \used as a baseline for new weights; it produced no GH or Le record.
\end{description}

\subsection{Verification and refinement}

\begin{description}[style=nextline,leftmargin=1.5em,itemsep=0.6ex]
\item[M\"oller's lower bound]
  A lower bound on the number of nodes for centrally symmetric weights, binding every rule,
  symmetric or not. A count that equals it is proven minimal.
  \origin \citet{molly79}, in the form stated by \citet{orive20}; background in
  \citet{xu25}.
  \used the ``M\"o'' column, and the only minimality statements in the tables.

\item[Relative gate in extended precision]
  Every rule is checked in BigFloat before it is banked: all weights must be positive, and
  the worst relative monomial error over all monomials of degree at most $p$,
  \[
    \max_{|a|\le p} \frac{|\sum_s w_s x_s^a - \mathbb{E}\,x^a|}{\max(\sum_s w_s\,|x_s^a|, 1)},
  \]
  must lie below $10^{-11}$. That is an acceptance threshold, not the accuracy achieved.
  The ``rel.\ err.''\ column of the tables is this error for the extended-precision file of
each rule as deposited, 80 significant digits for either weight, evaluated in arithmetic wide enough that rounding in the evaluation
plays no part. It is at most $7.2\cdot 10^{-70}$ for the uniform weight and
$9.3\cdot 10^{-69}$ for the Gaussian weight. The double-precision file of a rule is its
extended-precision file rounded to 17 digits.  Except for the trivial $p=1$ cases, its error
lies between $10^{-17}$ and $1.1\cdot 10^{-15}$ for the Gaussian weight and between $10^{-18}$
and $4\cdot 10^{-17}$ for the uniform weight, which is what rounding an exact rule to
double precision leaves.
  \origin this project.
  \used every cell.

\item[Extended-precision polish]
  An inexact Newton iteration with the residual in extended precision and a
  column-scaled Float64 Jacobian, reduced over $\pm$ pairs. A rule found under full $B_d$
  symmetry is a singular solution in that free-node frame, so such rules are polished in
  their orbit parameterization instead, where the system is small and regular and the whole
  Newton iteration runs in extended precision. One Gaussian rule ($d=4$, $p=9$), whose
  symmetry is not a union of complete $B_d$ orbits, has a Jacobian too nearly singular for
  either route and was refined with the Jacobian and the linear solve in extended precision
  as well. In every case the polish refines a rule; it never changes $N$. The deposited extended-precision files are also accurate in quadruple
  precision: with every node and weight rounded to IEEE binary128, the relative error above
  is at most $4.6$ machine epsilons ($2^{-112} \approx 1.9\cdot 10^{-34}$) for the Gaussian
  weight and $0.17$ for the uniform weight, the same multiples of the machine epsilon as in
  double precision. For the uniform weight the 80-digit files were in addition held to the
standard of \citetalias{diablo}, who stop their optimization at
$\|f - V^{\top}w\|_2 < 10^{-66}$, with $V$ the orthonormal Legendre product basis at the nodes on
$[-1,1]^d$ and weights summing to $2^d$. Evaluated in 512-bit arithmetic over every basis
product of degree at most $p$, and not only the fully symmetric ones that their rules
require, that norm lies between $10^{-78.8}$ and $10^{-66.5}$ in the 81 cells with $p>1$ and is
zero in the other four.
  \origin inexact Newton, \citet{dembo82}.
  \used the 80-digit files in the deposits.

\item[Monotone bank pass]
  A rule exact to degree $p+2$ is exact to degree $p$, so a smaller rule one degree up is
  copied down.
  \origin elementary.
  \used bookkeeping across every column.

\item[Rule-identity test]
  Two rules are compared with invariants that respect the symmetry of the weight: pairwise
  distance and (radius, weight) multisets for GH, which is rotation invariant, and the cube's
  own symmetry group for Le. The test decides whether a tie is the same rule or a different
  one.
  \origin this project.
  \used every ``same rule'' / ``different rule'' verdict in the comparisons.
\end{description}

\subsection{The counting floors in the tables}

\noindent
The fourth node-count column of each table --- pair in \cref{tab:best-le}, sym in
\cref{tab:best-gh} --- is a counting floor for a method rather than a bound for the
problem.

Every rule in \cref{tab:best-le} is centrally symmetric, so its nodes come in $±$pairs about the center of the cube and every moment of odd total degree is matched automatically.  Setting the $M_{\mathrm{even}}$ moment conditions that remain against the $d+1$ unknowns each pair contributes gives $N ≥ 2⌈M_{\mathrm{even}}/(d+1)⌉$, one node less where a node at the center helps.  The moment conditions need not be independent, however, and a rule carrying more symmetry can therefore sit below that count, which is why the column is a floor for the method rather than a bound for the problem.  The Gaussian rules do carry more symmetry than $±$pairing --- most of them are invariant under a sign change in each coordinate separately, and many under permutations of the coordinates as well --- and they fall below the pair count in 17 of their cells (not displayed).  The column sym in \cref{tab:best-gh} is the analogous count for symmetric rules.  A rule that is invariant under a group $G$ matches every polynomial that is not $G$-invariant automatically, so only the $G$-invariant moment conditions remain.  Its nodes come in orbits, and an orbit that lies in a symmetry plane has fewer nodes per unknown than a generic one, which is how symmetric rules get below the pair count.  For a given $G$, the count is the smallest number of nodes for which the unknowns cover the conditions, both overall and among the conditions that vanish on some orbit types and are hence seen by the other orbit types only; it is the solution of a small integer program over the orbit types.  The entry in the table is the minimum over $G$: no symmetry, $±$pairs, a sign change in each coordinate, the full symmetry group $B_d$ of the cube and, for $d=2$, the dihedral and cyclic groups.  Because the Gaussian weight is rotation invariant, the $d(d-1)/2$ unknowns that merely rotate a rule match no moment, and they are not counted where $G$ commutes with rotations; without that correction the count would fall below the Möller bound at low degrees.  No rule in \cref{tab:best-gh} has fewer nodes than sym, and 24 of the 57 have exactly that many, including all but four of those with $d=2$ and $p ≤ 27$.  What the two columns add is a sense of scale that the Möller bound loses as $d$ grows.  For the cube at $d=5$, $p=21$, the Möller bound is 3583 against my 10984 nodes, but the pair floor is 9820, so little room is plausibly left.  For the Gaussian weight in the same cell, my rule has 13199 nodes against a floor of 6290, which is the count for full $B_d$ symmetry, so there plausibly remains room.

\subsection{Tried without producing records}

\begin{description}[style=nextline,leftmargin=1.5em,itemsep=0.6ex]
\item[Orthogonal-polynomial ideal construction ($d=2$)]
  Nodes as common zeros of orthogonal polynomials, leaving a two-parameter family.
  \origin \citet{haggy76}.
  \used reproduced their Table 6 node for node, and showed that M\"oller's 24 at GH $d=2$,
  $p=11$ is not reachable within that family.

\item[Moment-matrix flat extension]
  Minimal cubature recast as rank minimization of a moment matrix.
  \origin \citet{curto96}.
  \used abandoned; it established a sharp trade-off between flatness and feasibility.

\item[Quantile-map warm starts between the two weights]
  A rule for one weight is carried to the other coordinate by coordinate, through the
  monotone map $F^{-1}\circ\Phi$ (or its inverse) or by matching the nodes of the
  one-dimensional Gauss rules, and the result is handed to the solver as a start.
  \origin the change of variables is standard; its use as a warm start is this project's.
  \used a negative result in both directions. From the Gaussian to the uniform weight the
  solver stalled at residuals between $10^{-3}$ and $0.5$, with the outer nodes pinned to
  the faces of the cube; from the uniform to the Gaussian weight ($d=2$, $p=31$--$39$) it
  converged from none of eleven starts. The mapped rules served only to suggest mixes of
  orbit types to the orbit searches.

\item[Product starts]
  The tensor product of a banked $(d-1)$-dimensional symmetric rule with the
  one-dimensional Gauss rule is an exact rule in $d$ dimensions; elimination then proceeds
  in the orbit space of the product.
  \origin product formulas are classical \citep{stroud71}; their use as a start for
  elimination is this project's.
  \used Le $d=5$, $p=15$--$21$. The products are far larger than the rules already banked
  ($21384$ nodes against $7200$ at $p=17$), and the descent never got below the bank.

\item[Weight homotopy from uniform to Gaussian]
  Continuation in a symmetric Jacobi weight $(1-x^2)^\alpha$ from $\alpha=0$ (uniform) toward
  the Gaussian limit, to carry cube rules over to $N(0,I_d)$.
  \origin this project.
  \used a negative result: every branch folds before reaching the Gaussian end.
\end{description}

\subsection{Which method produced the records}

\begin{center}
\setstretch{1}
\begin{tabular}{ll}
\hline
method & representative results \\
\hline
free-node and pair elimination & GH $d=2$, $p=17$--$33$; Le $d=2$ \\
$B_d$ orbit search & GH and Le, $d=3$--$5$, middle and high $p$ \\
$S_d\times\mathbb{Z}_2$ & GH $d=4$, $p=9$ ($116$); GH $d=5$, $p=9$ ($244$) \\
Diallo-Worku warm starts + pair elimination & Le $d=3$, $p=27$--$45$; Le $d=2$, $p=77$ \\
multithreaded pair elimination & Le $d=4$, $p=21, 23$; Le $d=5$, $p=17$--$21$ \\
degree continuation & GH $d=3$, $p=31, 33$ \\
\hline
\end{tabular}
\end{center}

\section*{Acknowledgments}

This research was done using services provided by the OSG Consortium \citep{osg07,osg09,ospool,osdf}, which is supported by the National Science Foundation awards \#2030508 and \#2323298.  The author recognizes the Penn State Institute for Computational and Data Sciences (ICDS) (RRID:SCR\_025154) for providing access to computational research infrastructure (RRID:SCR\_026424).

\section*{Artificial Intelligence}

Artificial intelligence (Claude) was used for the production of code and for managing the campaign across systems.  It was also used for literature searches, compiling the reference list, and to write much of \cref{app:methods}.  It was used to prepare and package up contents for submission to Zenodo. Finally, it was used to find typos and other infelicities in the text.

\begingroup
\footnotesize
\singlespacing
\setlength{\bibsep}{0pt}
\bibliographystyle{plainnat}
\bibliography{quadlit}

\begin{thebibliography}{45}
\providecommand{\natexlab}[1]{#1}
\providecommand{\url}[1]{\texttt{#1}}
\expandafter\ifx\csname urlstyle\endcsname\relax
  \providecommand{\doi}[1]{doi: #1}\else
  \providecommand{\doi}{doi: \begingroup \urlstyle{rm}\Url}\fi

\bibitem[Adurthi et~al.(2012)Adurthi, Singla, and Singh]{adurthi12}
Nagavenkat Adurthi, Puneet Singla, and Tarunraj Singh.
\newblock The conjugate unscented transform --- an approach to evaluate
  multi-dimensional expectation integrals.
\newblock In \emph{Proceedings of the American Control Conference}, 2012.
\newblock \doi{10.1109/ACC.2012.6314970}.

\bibitem[Bremer et~al.(2010)Bremer, Gimbutas, and Rokhlin]{bremer10}
James Bremer, Zydrunas Gimbutas, and Vladimir Rokhlin.
\newblock A nonlinear optimization procedure for generalized {G}aussian
  quadratures.
\newblock \emph{SIAM Journal on Scientific Computing}, 32\penalty0
  (4):\penalty0 1761--1788, 2010.
\newblock \doi{10.1137/080737046}.

\bibitem[Cools(2003)]{coolio}
Ronald Cools.
\newblock An encyclopaedia of cubature formulas.
\newblock \emph{Journal of Complexity}, 19:\penalty0 445--453, 2003.
\newblock \doi{10.1016/S0885-064X(03)00011-6}.
\newblock Online at \url{https://nines.cs.kuleuven.be/ecf/}.

\bibitem[Cools and Haegemans(1988)]{coolio88b}
Ronald Cools and Ann Haegemans.
\newblock Another step forward in searching for cubature formulae with a
  minimal number of knots for the square.
\newblock \emph{Computing}, 40:\penalty0 139--146, 1988.
\newblock \doi{10.1007/BF02247942}.

\bibitem[Curto and Fialkow(1996)]{curto96}
Ra{\'u}l~E. Curto and Lawrence~A. Fialkow.
\newblock Solution of the truncated complex moment problem for flat data.
\newblock \emph{Memoirs of the American Mathematical Society}, 119\penalty0
  (568), 1996.
\newblock \doi{10.1090/memo/0568}.

\bibitem[Degani et~al.(2005)Degani, Schiff, and Tannor]{degani05}
Ilan Degani, Jeremy Schiff, and David~J. Tannor.
\newblock Commuting extensions and cubature formulae.
\newblock \emph{Numerische Mathematik}, 101\penalty0 (3):\penalty0 479--500,
  2005.
\newblock \doi{10.1007/s00211-005-0628-z}.
\newblock Preprint arXiv:math/0408076. The degree-17 rule with 57 nodes for the
  Gaussian weight on the plane is in Section 5.

\bibitem[Dembo et~al.(1982)Dembo, Eisenstat, and Steihaug]{dembo82}
Ron~S. Dembo, Stanley~C. Eisenstat, and Trond Steihaug.
\newblock Inexact {N}ewton methods.
\newblock \emph{SIAM Journal on Numerical Analysis}, 19\penalty0 (2):\penalty0
  400--408, 1982.
\newblock \doi{10.1137/0719025}.

\bibitem[Diallo and Worku(2026)]{diablo}
Moustapha Diallo and Zelalem~Arega Worku.
\newblock High-order symmetric positive interior quadrature rules on two and
  three dimensional domains, 2026.
\newblock Preprint, arXiv:2601.14488 [math.NA]. Data:
  \url{https://github.com/mdiallo-fula/SymmetricPositiveInteriorCubatures.jl}.

\bibitem[Festa and Sommariva(2012)]{fiesta}
Mattia Festa and Alvise Sommariva.
\newblock Computing almost minimal formulas on the square.
\newblock \emph{Journal of Computational and Applied Mathematics}, 236\penalty0
  (17):\penalty0 4296--4302, 2012.
\newblock \doi{10.1016/j.cam.2012.05.021}.

\bibitem[Frontin et~al.(2021)Frontin, Walters, Witherden, Lee, Williams, and
  Darmofal]{frontin21}
Cory~V. Frontin, Gage~S. Walters, Freddie~D. Witherden, Carl~W. Lee, David~M.
  Williams, and David~L. Darmofal.
\newblock Foundations of space-time finite element methods: polytopes,
  interpolation, and integration.
\newblock \emph{Applied Numerical Mathematics}, 166:\penalty0 92--113, 2021.
\newblock \doi{10.1016/j.apnum.2021.03.019}.

\bibitem[Genz and Malik(1980)]{genz80}
A.~C. Genz and A.~A. Malik.
\newblock Remarks on algorithm 006: An adaptive algorithm for numerical
  integration over an {N}-dimensional rectangular region.
\newblock \emph{Journal of Computational and Applied Mathematics}, 6\penalty0
  (4):\penalty0 295--302, 1980.
\newblock \doi{10.1016/0771-050X(80)90039-X}.

\bibitem[Genz and Keister(1996)]{genz96}
Alan Genz and B.~D. Keister.
\newblock Fully symmetric interpolatory rules for multiple integrals over
  infinite regions with {G}aussian weight.
\newblock \emph{Journal of Computational and Applied Mathematics}, 71\penalty0
  (2):\penalty0 299--309, 1996.
\newblock \doi{10.1016/0377-0427(95)00232-4}.

\bibitem[Glaubitz(2023)]{glaubitz21}
Jan Glaubitz.
\newblock Construction and application of provable positive and exact cubature
  formulas.
\newblock \emph{IMA Journal of Numerical Analysis}, 43\penalty0 (3):\penalty0
  1616--1652, 2023.
\newblock \doi{10.1093/imanum/drac017}.

\bibitem[Golub and Welsch(1969)]{golub69}
Gene~H. Golub and John~H. Welsch.
\newblock Calculation of {G}auss quadrature rules.
\newblock \emph{Mathematics of Computation}, 23\penalty0 (106):\penalty0
  221--230, 1969.
\newblock \doi{10.1090/S0025-5718-69-99647-1}.

\bibitem[Haegemans and Piessens(1976)]{haggy76}
Ann Haegemans and Robert Piessens.
\newblock Construction of cubature formulas of degree eleven for symmetric
  planar regions, using orthogonal polynomials.
\newblock \emph{Numerische Mathematik}, 25:\penalty0 139--148, 1976.
\newblock \doi{10.1007/BF01462267}.

\bibitem[Haegemans and Piessens(1977)]{haggy77}
Ann Haegemans and Robert Piessens.
\newblock Construction of cubature formulas of degree seven and nine for
  symmetric planar regions, using orthogonal polynomials.
\newblock \emph{SIAM Journal on Numerical Analysis}, 14\penalty0 (3):\penalty0
  492--508, 1977.
\newblock \doi{10.1137/0714029}.

\bibitem[Heiss and Winschel(2008)]{heiss08}
Florian Heiss and Viktor Winschel.
\newblock Likelihood approximation by numerical integration on sparse grids.
\newblock \emph{Journal of Econometrics}, 144\penalty0 (1):\penalty0 62--80,
  2008.
\newblock \doi{10.1016/j.jeconom.2007.12.004}.

\bibitem[Keshavarzzadeh et~al.(2018)Keshavarzzadeh, Kirby, and Narayan]{cash}
Vahid Keshavarzzadeh, Robert~M. Kirby, and Akil Narayan.
\newblock Numerical integration in multiple dimensions with designed
  quadrature.
\newblock \emph{SIAM Journal on Scientific Computing}, 40\penalty0
  (4):\penalty0 A2033--A2061, 2018.
\newblock \doi{10.1137/17M1137875}.

\bibitem[Konyaev(1977)]{konyaev77}
S.~I. Konyaev.
\newblock Kvadraturnye formuly 9-go poryadka, invariantnye otnositel'no gruppy
  ikosa\`edra [{N}inth-order quadrature formulas invariant with respect to the
  icosahedral group].
\newblock \emph{Doklady Akademii Nauk SSSR}, 233\penalty0 (5):\penalty0
  784--787, 1977.
\newblock In Russian.

\bibitem[Lawson and Hanson(1974)]{lawson74}
Charles~L. Lawson and Richard~J. Hanson.
\newblock \emph{Solving Least Squares Problems}.
\newblock Prentice-Hall, Englewood Cliffs, NJ, 1974.

\bibitem[Levenberg(1944)]{levenberg44}
Kenneth Levenberg.
\newblock A method for the solution of certain non-linear problems in least
  squares.
\newblock \emph{Quarterly of Applied Mathematics}, 2:\penalty0 164--168, 1944.
\newblock \doi{10.1090/qam/10666}.

\bibitem[Liu and Pierce(1994)]{liu94}
Qing Liu and Donald~A. Pierce.
\newblock A note on {G}auss--{H}ermite quadrature.
\newblock \emph{Biometrika}, 81\penalty0 (3):\penalty0 624--629, 1994.
\newblock \doi{10.1093/biomet/81.3.624}.

\bibitem[Madsen et~al.(2004)Madsen, Nielsen, and Tingleff]{madsen04}
K.~Madsen, H.~B. Nielsen, and O.~Tingleff.
\newblock \emph{Methods for Non-Linear Least Squares Problems}.
\newblock Informatics and Mathematical Modelling, Technical University of
  Denmark, 2nd edition, 2004.

\bibitem[Marquardt(1963)]{marquardt63}
Donald~W. Marquardt.
\newblock An algorithm for least-squares estimation of nonlinear parameters.
\newblock \emph{Journal of the Society for Industrial and Applied Mathematics},
  11\penalty0 (2):\penalty0 431--441, 1963.
\newblock \doi{10.1137/0111030}.

\bibitem[Molien(1897)]{molien97}
Theodor Molien.
\newblock {\"U}ber die {I}nvarianten der linearen {S}ubstitutionsgruppen.
\newblock \emph{Sitzungsberichte der K{\"o}niglich Preu{\ss}ischen Akademie der
  Wissenschaften zu Berlin}, pages 1152--1156, 1897.

\bibitem[M\"{o}ller(1976)]{molly76}
H.~M. M\"{o}ller.
\newblock Kubaturformeln mit minimaler {K}notenzahl.
\newblock \emph{Numerische Mathematik}, 25:\penalty0 185--200, 1976.
\newblock \doi{10.1007/BF01462272}.

\bibitem[M\"{o}ller(1979)]{molly79}
H.~M. M\"{o}ller.
\newblock Lower bounds for the number of nodes in cubature formulae.
\newblock In \emph{Numerische Integration}, volume~45 of \emph{ISNM}, pages
  221--230. Birkh\"{a}user, 1979.
\newblock \doi{10.1007/978-3-0348-6288-2_17}.

\bibitem[Naylor and Smith(1982)]{naylor82}
J.~C. Naylor and A.~F.~M. Smith.
\newblock Applications of a method for the efficient computation of posterior
  distributions.
\newblock \emph{Journal of the Royal Statistical Society, Series C (Applied
  Statistics)}, 31\penalty0 (3):\penalty0 214--225, 1982.
\newblock \doi{10.2307/2347995}.

\bibitem[Nielsen(1999)]{nielsen99}
Hans~Bruun Nielsen.
\newblock Damping parameter in {M}arquardt's method.
\newblock Technical Report IMM-REP-1999-05, Informatics and Mathematical
  Modelling, Technical University of Denmark, 1999.

\bibitem[Orive et~al.(2020)Orive, Santos-Le\'{o}n, and Spalevi\'{c}]{orive20}
Ram\'{o}n Orive, Juan~C. Santos-Le\'{o}n, and Miodrag~M. Spalevi\'{c}.
\newblock Cubature formulae for the {G}aussian weight. {S}ome old and new
  rules.
\newblock \emph{Electronic Transactions on Numerical Analysis}, 53:\penalty0
  426--438, 2020.
\newblock \doi{10.1553/etna_vol53s426}.

\bibitem[{OSG}(2006)]{ospool}
{OSG}.
\newblock {OSPool}, 2006.
\newblock URL \url{https://osg-htc.org/services/open_science_pool.html}.

\bibitem[{OSG}(2015)]{osdf}
{OSG}.
\newblock {O}pen {S}cience {D}ata {F}ederation, 2015.
\newblock URL \url{https://osdf.osg-htc.org/}.

\bibitem[Pordes et~al.(2007)Pordes, Petravick, Kramer, Olson, Livny, Roy,
  Avery, Blackburn, Wenaus, W{\"u}rthwein, Foster, Gardner, Wilde, Blatecky,
  McGee, and Quick]{osg07}
Ruth Pordes, Don Petravick, Bill Kramer, Doug Olson, Miron Livny, Alain Roy,
  Paul Avery, Kent Blackburn, Torre Wenaus, Frank W{\"u}rthwein, Ian Foster,
  Rob Gardner, Mike Wilde, Alan Blatecky, John McGee, and Rob Quick.
\newblock The open science grid.
\newblock \emph{Journal of Physics: Conference Series}, 78:\penalty0 012057,
  2007.
\newblock \doi{10.1088/1742-6596/78/1/012057}.

\bibitem[Sfiligoi et~al.(2009)Sfiligoi, Bradley, Holzman, Mhashilkar, Padhi,
  and W{\"u}rthwein]{osg09}
Igor Sfiligoi, Daniel~C. Bradley, Burt Holzman, Parag Mhashilkar, Sanjay Padhi,
  and Frank W{\"u}rthwein.
\newblock The pilot way to grid resources using {glideinWMS}.
\newblock In \emph{2009 WRI World Congress on Computer Science and Information
  Engineering}, volume~2, pages 428--432, 2009.
\newblock \doi{10.1109/CSIE.2009.950}.

\bibitem[Slobodkins and Tausch(2023)]{slobodkins23}
Arkadijs Slobodkins and Johannes Tausch.
\newblock A node elimination algorithm for cubature of high-dimensional
  polytopes.
\newblock \emph{Computers \& Mathematics with Applications}, 144:\penalty0
  229--236, 2023.
\newblock \doi{10.1016/j.camwa.2023.06.001}.

\bibitem[Smolyak(1963)]{smolyak63}
S.~A. Smolyak.
\newblock Quadrature and interpolation formulas for tensor products of certain
  classes of functions.
\newblock \emph{Soviet Mathematics Doklady}, 4:\penalty0 240--243, 1963.
\newblock Russian original: Doklady Akademii Nauk SSSR 148 (1963) 1042--1045.

\bibitem[Sobolev(1962)]{sobolev62}
S.~L. Sobolev.
\newblock Cubature formulas on the sphere invariant under finite groups of
  rotations.
\newblock \emph{Doklady Akademii Nauk SSSR}, 146:\penalty0 310--313, 1962.

\bibitem[Stroud(1971)]{stroud71}
A.~H. Stroud.
\newblock \emph{Approximate Calculation of Multiple Integrals}.
\newblock Prentice-Hall, Englewood Cliffs, NJ, 1971.

\bibitem[Stroud and Secrest(1963)]{stroud63}
A.~H. Stroud and Don Secrest.
\newblock Approximate integration formulas for certain spherically symmetric
  regions.
\newblock \emph{Mathematics of Computation}, 17:\penalty0 105--135, 1963.
\newblock \doi{10.1090/S0025-5718-1963-0161473-0}.

\bibitem[Tchakaloff(1957)]{tchakaloff57}
Vladimir Tchakaloff.
\newblock Formules de cubatures m{\'e}caniques {\`a} coefficients non
  n{\'e}gatifs.
\newblock \emph{Bulletin des Sciences Math{\'e}matiques}, 81:\penalty0
  123--134, 1957.

\bibitem[Transtrum and Sethna(2012)]{transtrum12}
Mark~K. Transtrum and James~P. Sethna.
\newblock Improvements to the {L}evenberg--{M}arquardt algorithm for nonlinear
  least-squares minimization, 2012.
\newblock arXiv:1201.5885.

\bibitem[Van~Zandt(2019)]{sandy}
James~R. Van~Zandt.
\newblock Efficient cubature rules.
\newblock \emph{Electronic Transactions on Numerical Analysis}, 51:\penalty0
  219--239, 2019.
\newblock \doi{10.1553/etna_vol51s219}.

\bibitem[Vioreanu and Rokhlin(2014)]{vioreanu14}
Bogdan Vioreanu and Vladimir Rokhlin.
\newblock Spectra of multiplication operators as a numerical tool.
\newblock \emph{SIAM Journal on Scientific Computing}, 36\penalty0
  (1):\penalty0 A267--A288, 2014.
\newblock \doi{10.1137/110860082}.

\bibitem[Xiao and Gimbutas(2010)]{xiao10}
Hong Xiao and Zydrunas Gimbutas.
\newblock A numerical algorithm for the construction of efficient quadrature
  rules in two and higher dimensions.
\newblock \emph{Computers \& Mathematics with Applications}, 59:\penalty0
  663--676, 2010.
\newblock \doi{10.1016/j.camwa.2009.10.027}.

\bibitem[Xu(2025)]{xu25}
Yuan Xu.
\newblock \emph{Minimal Cubature Rules: Theory and Practice}.
\newblock Cambridge Monographs on Applied and Computational Mathematics.
  Cambridge University Press, 2025.
\newblock ISBN 9781009663939.
\newblock \doi{10.1017/9781009663939}.

\end{thebibliography}
\endgroup

\end{document}